\documentclass[superscriptaddress,twocolumn,amsmath,amssymb]{revtex4}
\usepackage{graphicx}
\usepackage{color}

\renewcommand{\vec}[1]{\mbox{\boldmath $#1$}}

\begin{document}

\title{New and efficient method for solving the eigenvalue problem 
for the 
two-center shell model with finite-depth potentials} 

\author{K. Hagino}
\affiliation{
Department of Physics, Tohoku University, Sendai 980-8578,  Japan}
\affiliation{Research Center for Electron Photon Science, Tohoku
University, 1-2-1 Mikamine, Sendai 982-0826, Japan}
\affiliation{
National Astronomical Observatory of Japan, 2-21-1 Osawa,
Mitaka, Tokyo 181-8588, Japan}

\author{T. Ichikawa}
\affiliation{
Center for Nuclear Study, University of Tokyo, Tokyo 113-0033, Japan}


\begin{abstract}
We propose a new method to solve the eigen-value problem with 
a two-center single-particle potential. 
This method combines 
the usual matrix diagonalization with 
the method of separable representation of 
a two-center potential, that 
is, an expansion of the two-center potential with a finite basis set. 
To this end, we expand the potential on a harmonic oscillator basis, 
while single-particle wave functions on a combined basis with a harmonic 
oscillator and eigen-functions of a one-dimensional two-center potential. 
In order to demonstrate its efficiency, we apply this method to a system with 
two $^{16}$O nuclei, in which the potential is given as a sum of two 
Woods-Saxon potentials. 
\end{abstract}

\maketitle

\section{Introduction}

A single-particle motion in a two-center potential 
\cite{GPS94,MG72,Gherghescu03} 
is an important 
ingredient in understanding the dynamics of heavy-ion fusion reactions 
and nuclear fission \cite{GM74,SW75,IM13,Diaz-Torres04,Diaz-Torres07}. 
In particular, 
a Landau-Zener transition at level 
crossing points plays an important role in dissipative phenomena 
in the nuclear dynamics \cite{GM74,SW75}. 
Single-particle levels in a two-center potential also provide a 
basis to calculate the shell correction energy in a potential energy 
surface for fission \cite{ICA14} as well as 
for fusion to synthesize superheavy elements \cite{ZG15}.  

In the past, the two-center shell model has been solved 
with various methods. These are mainly categorized into two approaches. 
The first approach is to expand single-particle wave functions on some basis 
and then to obtain eigen-functions by diagonalizing 
the Hamiltonian matrix. For this purpose, the two-center harmonic oscillator 
basis \cite{HMG69}, a deformed harmonic oscillator basis 
with single-center \cite{IM13,ZM76,PL77}, and a non-orthogonal 
two-center basis \cite{Hasse74,BS74,OS75,NSP87} have been used. 
The second approach, on the other hand, is to expand each potential in a two-center potential 
on some basis and then to shift it with a quantum mechanical 
shift operator \cite{GGR77,MR86}. 
To obtain eigen-functions for the resultant potential, 
the single-particle Schr\"odinger equation is transformed to a linear 
algebraic equation based on the 
Lippmann-Schwinger equation,  
and then the eigen-values are sought by checking 
the solvability condition of the equation as a function of a 
single-particle energy \cite{GGR77,MR86,GKR79,Diaz-Torres05,Diaz-Torres08}. 

Each approach has both advantages and disadvantages. 
For the matrix diagonalization method, the method itself 
is conceptually simple and one can apply it easily even when two 
single-particle energies are nearly degenerate in energy at a level crossing point. 
A disadvantage of this method, however, 
is that it is not easy to obtain 
an efficient basis to represent single-particle wave functions. 
The two-center oscillator basis is efficient, but this 
basis involves confluent hypergeometric functions and thus it may not be 
easy to construct the basis. The deformed oscillator 
basis is straightforward to use, but a large number of basis states is 
required at large separation distances of 
two potential wells. This problem can be avoided by 
using the non-orthogonal two-center basis, but calculations with 
such basis may suffer from a numerical instability at short distances due to 
the overcompleteness of the basis \cite{TRD77}. 
Moreover, with these basis functions, it is not straightforward 
to compute matrix elements 
of a spin-orbit potential in single-particle potentials when they are 
shifted from the origin. 

In contrast, a spin-orbit potential is easily evaluated with the 
second approach, at least when the potential is spherical, 
since with this approach one first calculates the matrix elements of a potential centered at 
the origin. Also, the linear algebraic equations may be solved easily due to its simple structure 
originated from the separable representation of a two-center potential. 
A disadvantage of this approach, however, 
is that a care must be taken when two single-particle 
energies are close to each other in seeking the solvability condition 
of the equation. One also has to use different treatments for bound 
states and scattering states because of the different boundary conditions of 
the wave functions \cite{MR86}. 
Another point is that the matrix elements of a Green's function have to 
be constructed at each energy, which may be time consuming if many basis 
states are included in a calculation, even though one may be able to resort to a 
recurrence formula \cite{MR86,GKR79}. 

In this paper, we propose a novel method for the two-center shell model, 
which combines good aspects of the previous two approaches. In this new method, 
we directly diagonalize a single-particle Hamiltonian, in which  
a two-center potential is expanded on a harmonic oscillator basis  
as in the second method. In this way, a spin-orbit potential can 
be evaluated in a straightforward manner. Also, by diagonalizing a Hamiltonian matrix, 
one can easily 
obtain eigen-functions even at a level crossing point, as in the first 
method. 
A similar method has been employed in Ref. \cite{Diaz-Torres04-2}, but 
for a single-center potential.
In this paper, for simplicity, we consider two spherical 
single-particle potentials shifted at two different positions, so that 
the resultant two-center potential has an axially symmetric shape. 
In order to expand single-particle wave functions, we then employ 
a harmonic oscillator basis for the direction perpendicular to the 
symmetric axis while we use eigen-functions of a one-dimensional 
single-particle two-center 
potential well as a basis for the direction along the symmetric axis. 
Such basis is efficient both at large and small separation 
distances, and yet it is 
easy to handle in evaluating several matrix elements. 

The paper is organized as follows. In Sec. II, we formulate our 
new method for the two-center shell model. We apply the method in 
Sec. III to a system with two $^{16}$O nuclei. To this end, 
we use a two-center 
single-particle potential with two shifted spherical Woods-Saxon 
potentials. We shall compare the results with calculations with a harmonic oscillator 
basis and discuss the efficiency of our method. We finally summarize 
the paper in Sec. IV. 

\section{New approach to Two-center shell model}

\subsection{General formalism}

\begin{figure}[t]
\includegraphics[clip,width=7cm]{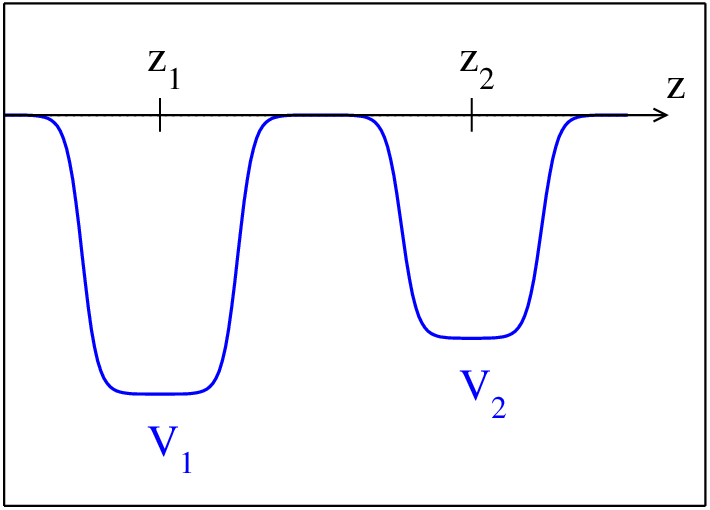}
\caption{A schematic view of a two-center potential given by 
Eq. (\ref{pot}). }
\end{figure}

We consider a single-particle motion of a particle with mass $m$ in a 
potential which consists of two 
potential wells located at $z_1$ and $z_2$ on the $z$-axis (see Fig. 1): 
\begin{equation}
V(\vec{r})=V_1(\vec{r}-z_1\vec{e}_z)+V_2(\vec{r}-z_2\vec{e}_z),
\label{pot}
\end{equation}
where $\vec{e}_z$ is the unit vector in the $z$-direction. 
We first notice that the shifted potentials can be expressed as 
\cite{GGR77,MR86}, 
\begin{equation}
V_s(\vec{r}-z_s\vec{e}_z)=e^{-i\hat{p}_zz_s/\hbar}\,V_s(\vec{r})\,
e^{i\hat{p}_zz_s/\hbar}~~~~(s=1,2),
\label{pot2}
\end{equation}
where $\hat{p}_z$
is the usual momentum operator for the $z$-direction. 
The idea of the separable expansion method \cite{GKR79,MR86,GGR77} 
is to expand the potentials 
$V_s$ on some basis as, 
\begin{eqnarray}
V_s(\vec{r})&=&\sum_{\alpha,\alpha'}\,|\Psi_\alpha\rangle
\langle \Psi_\alpha|V_s|\Psi_{\alpha'}\rangle
\langle \Psi_{\alpha'}|, \\
&\equiv& 
\sum_{\alpha,\alpha'}\,|\Psi_\alpha\rangle \,V_{\alpha\alpha'}^{(s)}\,
\langle \Psi_{\alpha'}|, 
\end{eqnarray}
where $\{|\Psi_\alpha\rangle\}$ is a set of the basis functions 
(one could use different basis sets between $V_1$ and $V_2$, but here 
we use the same basis in order to simplify the notation). 
The single-particle potential, Eq. (\ref{pot}), then reads,
\begin{equation}
V=\sum_{s=1,2}
\sum_{\alpha,\alpha'}\,e^{-i\hat{p}_zz_s/\hbar}\,|\Psi_\alpha\rangle \,V_{\alpha\alpha'}^{(s)}
\,\langle \Psi_{\alpha'}|\,e^{i\hat{p}_zz_s/\hbar}.
\label{pot3}
\end{equation}

In Refs. \cite{GKR79,MR86,GGR77}, the Schr\"odinger equation with 
the potential given by Eq. (\ref{pot3}), that is, 
\begin{equation}
\left(\frac{\vec{p}^2}{2m}+V(\vec{r})-E\right)\Psi(\vec{r})=0,
\label{Schrodinger}
\end{equation}
where $\Psi$ is a single-particle wave 
function, 
is first transformed to the Lippmann-Schwinger equation. 
For a bound state, it reads, 
\begin{equation}
|\Psi\rangle = \frac{1}{E-\frac{\vec{p}^2}{2m}}\,V|\Psi\rangle = G_0(E)\,V|\Psi\rangle, 
\end{equation}
where 
\begin{equation}
G_0(E)=\frac{1}{E-\frac{\vec{p}^2}{2m}},
\end{equation}
is the Green's function. 
From this equation, one obtains, 
\begin{eqnarray}
&& \sum_{s'}\sum_\beta\left[
\delta_{s,s'}\delta_{\alpha,\beta}\right. \nonumber \\
&& \left.-\sum_{\alpha'}
\langle\Psi_\alpha|G_0(E)e^{i\hat{p}_z(z_s-z_{s'})}|\Psi_{\alpha'}\rangle
V_{{\alpha'} \beta}^{(s')}
\right]x_{\beta s'}=0,
\label{separable}
\end{eqnarray}
with $x_{\alpha s}\equiv \langle \Psi_\alpha|e^{i\hat{p}_zz_s/\hbar}|\Psi\rangle$. 
The eigen-values $E$ can be found by requiring that the 
determinant of the matrix in Eq. (\ref{separable}) 
vanishes at $E$ \cite{GKR79,MR86,GGR77}. 

This method has been employed in several applications in the past. 
For instance, the author of Ref. \cite{Diaz-Torres08} used 
this method to discusses the two-center problem with arbitrarily 
oriented deformed potentials. It was also applied in Ref. \cite{MR86} 
to a problem of nucleon emission in 
heavy-ion collisions.  
However, as we have mentioned in the previous section, this method may have 
a difficulty when two eigen-energies are close to each 
other. 

We therefore attempt to solve directly 
the Schr\"odinger equation, Eq. (\ref{Schrodinger}), with the separable 
representation of the single-particle potential, Eq. (\ref{pot3}). 
To this end, we expand the single-particle wave function, $\Psi$, 
on a basis as, 
\begin{equation}
|\Psi\rangle = \sum_k\,C_k|\tilde{\Psi}_k\rangle, 
\end{equation}
where the basis set $\{|\tilde{\Psi}_k\rangle\}$ is in general different 
from the basis set $\{|\Psi_\alpha\rangle\}$ for the 
potential. 
Using Eq. (\ref{pot3}), one then obtains 
\begin{equation}
\sum_{k,k'}H_{kk'}C_{k'}=E\,C_k, 
\end{equation}
with $H_{kk'}=T_{kk'}+V_{kk'}$, where $T_{kk'}$ and $V_{kk'}$ are given by, 
\begin{equation}
T_{kk'}=\left\langle\tilde{\Psi}_k\left| 
\frac{\vec{p}^2}{2m}\right|\tilde{\Psi}_{k'}\right\rangle,
\label{T}
\end{equation}
and
\begin{equation}
V_{kk'}=\sum_{s=1,2}
\sum_{\alpha,\alpha'}\,
\langle\tilde{\Psi}_k|e^{-i\hat{p}_zz_s/\hbar}\,|\Psi_\alpha\rangle \,
V_{\alpha\alpha'}^{(s)}
\,\langle \Psi_{\alpha'}|\,e^{i\hat{p}_zz_s/\hbar}|\tilde{\Psi}_{k'}\rangle,
\label{V}
\end{equation}
respectively. 
The eigen-values and the eigen-functions are obtained by numerically 
diagonalizing the Hamiltonian matrix, $\{H_{kk'}\}$. 
Notice that they can be obtained at once in this method, both for 
bound and continuum states, once the Hamiltonian 
matrix is constructed, 
whereas the matrix elements need to be constructed at each $E$ in the 
previous method. 
Note also that this method can easily be applied even in a situation 
when two eigen-values are close to each other. 
In general, one obtains both negative and positive energy 
states by the diagonalizing procedure. 
The positive wave functions 
so obtained well represent the inner part of scattering 
wave function at the same energy, even though the outer part reflects 
the properties of the basis functions and thus may not be well 
described. See e.g., Ref. \cite{HT70}. 

\subsection{Harmonic oscillator basis}

In this paper, we consider a spherical central 
potential, together with a spin-orbit potential, for each of the potential 
wells, $V_s$. 
To be more specific, we consider a potential in a form of 
\begin{equation}
V_s(r)=V^{(s)}_0(r)+V^{(s)}_{\rm ls}(r)\,\vec{l}\cdot\vec{s}, 
\label{Vsphe}
\end{equation}
where $\vec{l}=\vec{r}\times\vec{p}/\hbar$ 
and $\vec{s}$ are the orbital and the spin angular momenta, 
respectively, and $V^{(s)}_0(r)$ and $V^{(s)}_{\rm ls}(r)$ are assumed to depend only on 
$r=|\vec{r}|$. For this problem, 
we particularly employ a harmonic oscillator basis 
in the general formalism presented in the previous subsection. 
In the cylindrical coordinate, this basis is given by \cite{Vautherin73}, 
\begin{equation}
\Psi_\alpha(\vec{r})=\Psi_{n_z n_\rho \Lambda m_s}(\vec{r})
=\phi_{n_z}(z)\,\psi_{n_\rho}^{(\Lambda)}(\rho)\,
\frac{e^{i\Lambda \phi}}{\sqrt{2\pi}}\,\chi_{m_s},
\label{hobasis}
\end{equation}
where $\rho=\sqrt{x^2+y^2}$ with $x=\rho\cos\phi$ and $y=\rho\sin\phi$. 
Here, $\chi_{m_s}$ is the spin wave function, $m_s$ and $\Lambda$ being the 
$z$-component of the spin and the orbital angular momenta, respectively. 
The functions $\phi_{n_z}(z)$ and $\psi_{n_\rho}^{(\Lambda)}(\rho)$ in 
Eq. (\ref{hobasis}) are given by, 
\begin{eqnarray}
\phi_{n_z}(z)&=&\sqrt{\frac{1}{\sqrt{\pi}2^{n_z}n_z!}}\,b_0^{-1/2}\,
e^{-\frac{z^2}{2b_0^2}}\,H_{n_z}(z/b_0), 
\label{HO_z}
\\
\psi_{n_\rho}^{(\Lambda)}(\rho)&=&
\sqrt{\frac{n_\rho!}{(n_\rho+\Lambda)!}}\,\frac{\sqrt{2}}{b_0}\,
\left(\frac{\rho}{b_0}\right)^\Lambda\,e^{-\frac{\rho^2}{2b_0^2}}\,
L_{n_\rho}^{(\Lambda)}(\rho^2/b_0^2), \nonumber \\
\end{eqnarray}
respectively, where $H_{n_z}$ and $L_{n_\rho}^{(\Lambda)}$ 
are the Hermite polynomials and the associated Laguerre polynomials, 
respectively. $b_0=\sqrt{\hbar/(m\omega_0)}$ is the oscillator 
length, with which the basis function satisfies the equation, 
\begin{equation}
\left[\frac{\vec{p}^2}{2m}+\frac{1}{2}m\omega_0^2\,(\rho^2+z^2)
-\epsilon_{n_z n_\rho \Lambda}\right]
\Psi_{n_z n_\rho \Lambda m_s}(\vec{r})=0,
\end{equation}
with $\epsilon_{n_z n_\rho \Lambda}=(n_z+2n_\rho+\Lambda+3/2)\,\hbar\omega_0$. 
Notice that we employ the same oscillator length for the $z$ and $\rho$ 
directions, since the potentials, Eq. (\ref{Vsphe}), are both spherical. 
The matrix elements of the potentials with this basis are given in 
Appendix. 

For the basis for the wave functions, $\tilde{\Psi}_k$, we use the 
same harmonic oscillator basis as in Eq. (\ref{hobasis}) for the $\rho$-direction (with the same 
oscillator length, $b_0$) while we  
use a different function, $\tilde{\phi}_{\tilde{n}_z}(z)$, which is 
to be specific below, for the $z$-direction. 
That is, the basis for the wave functions reads, 
\begin{equation}
\tilde{\Psi}_k(\vec{r})=
\tilde{\Psi}_{\tilde{n}_z \tilde{n}_\rho \tilde{\Lambda} \tilde{m}_s}(\vec{r})
=\tilde{\phi}_{\tilde{n}_z}(z)\,\psi_{\tilde{n}_\rho}^{(\tilde{\Lambda})}(\rho)\,
\frac{e^{i\tilde{\Lambda} \phi}}{\sqrt{2\pi}}\,\chi_{\tilde{m}_s}.
\label{hobasis_wf}
\end{equation}
The overlap integrals in the matrix elements for 
the single-particle potentials, Eq. (\ref{V}), 
are then given by, 
\begin{eqnarray}
\langle\tilde{\Psi}_k|e^{-i\hat{p}_zz_s/\hbar}\,|\Psi_\alpha\rangle 
&=&\delta_{n_\rho,\tilde{n}_\rho} \delta_{\Lambda,\tilde{\Lambda}} 
\delta_{m_s,\tilde{m}_s}\, \nonumber \\
&& \times \int^\infty_{-\infty}dz\, \tilde{\phi}_{\tilde{n}_z}(z) 
\phi_{n_z}(z-z_s), \nonumber \\
\end{eqnarray}
where we have assumed that the basis function $\tilde{\phi}_n(z)$ is a 
real function of $z$. 
If one takes the harmonic oscillator basis, Eq. (\ref{HO_z}), 
for $\tilde{\phi}_n(z)$ (but with a different oscillator length from $b_0$), 
these overlap integrals can be computed 
analytically \cite{DW86,II98,Chang05,GME06}. 
Instead, we here use the eigen-functions for the one-dimensional 
central potential,  
\begin{equation}
V_z(z)=V_0^{(1)}(|z-z_1|)+V_0^{(2)}(|z-z_2|),
\end{equation}
which satisfy the one-dimensional Schr\"odinger equation of, 
\begin{equation}
\left(-\frac{\hbar^2}{2m}\,\frac{d^2}{dz^2}+V_z(z)-\epsilon_z\right)
\tilde{\phi}_{\tilde{n}_z}(z)=0. 
\label{eigen-z}
\end{equation}
Here, we use only the central part of the three-dimensional potential, 
Eq. (\ref{Vsphe}). 
We solve this equation numerically with the Numerov method \cite{Koonin} 
in order to obtain the basis functions, $\tilde{\phi}_{\tilde{n}_z}(z)$.
The continuum states may be discretized by imposing 
the box boundary condition at $z=\pm z_{\rm box}$. 

\section{Application to $^{16}$O+$^{16}$O system}

We now apply the new method for the two-center shell model to an actual 
problem. For this purpose, we consider neutron single-particle states 
in the $^{16}$O+$^{16}$O system \cite{Diaz-Torres07,ZM76,GGR77}. 
The two potential wells, Eq. (\ref{Vsphe}), 
are then identical to each other, that is, $V_1(r)=V_2(r)$. 
For these potential wells, we employ the Woods-Saxon form, that is, 
\begin{eqnarray}
V_1(r)=V_2(r)&=&V_0\left(1-\kappa(\vec{l}\cdot\vec{s})
\frac{1}{r}\frac{d}{dr}\right)\nonumber \\
&&~~~\times \left[1+\exp\left(\frac{r-R}{a}\right)\right]^{-1}. 
\label{eq:WS}
\end{eqnarray}
We use the same values for the parameters as those in 
Ref. \cite{MR86}, that is, $V_0=-50.2$ MeV, $R=1.24\times 16^{1/3}$ fm, 
$\kappa=0.524$ fm$^2$,  
and $a=0.63$ fm, together with $m=939.6$ MeV/$c^2$. 

We take into account the volume conservation condition 
in a similar way as in Refs. \cite{Diaz-Torres05,NSP87}. 
That is, the parameters $V_0$ and $R$ in the Woods-Saxon potential, Eq. (\ref{eq:WS}), 
are adjusted 
at each separation distance, $z=|z_1-z_2|$,  
so that the two-center potential given by Eq. (\ref{pot}) 
is smoothly connected to a one-center potential for the 
$^{32}$S nucleus as the separation distance 
is decreased to zero. 
For this purpose, we assume 
the same Woods-Saxon potential for $^{32}$S as in Eq. (\ref{eq:WS}) with 
the same value of $\kappa$ and $a$, while the radius parameter is modified to 
$R=1.24\times 32^{1/3}$ fm. For this potential, the depth parameter, $V_0$, is slightly 
adjusted to be $V_0=-51.823$ MeV so that the volume conservation condition 
is satisfied (see below). 
In order to determine the value of the parameters at intermediate distances, 
we interpolate the parameters $V_0$ and $1/R$ between those at $z=0$ 
and $z=\infty$ \cite{Diaz-Torres05,NSP87}, that is, 
\begin{eqnarray}
V_0&=&(1-x)\, V_0(^{32}{\rm S}) +x \,V_0(^{16}{\rm O}), \\
\frac{1}{R}&=&(1-x)\,\frac{1}{R(^{32}{\rm S})}+x\,\frac{1}{R(^{16}{\rm O})}, 
\end{eqnarray} 
where $V_0(^{32}{\rm S})$ and $R(^{32}{\rm S})$ are the depth and the radius parameters for the $^{32}$S nucleus, 
respectively, while $V_0(^{16}{\rm O})$ and $R(^{16}{\rm O})$ are those for the $^{16}$O nucleus. 
The value of $x$ is determined by imposing 
the 
volume conservation condition given by, 
\begin{equation}
\int d\vec{r}\,\theta\left[W_0-\tilde{V}_0(\vec{r};z_1,z_2)
\right] \,\tilde{V}_0(\vec{r};z_1,z_2)= {\rm const.}, 
\label{eq:volume}
\end{equation}
with 
\begin{equation}
\tilde{V}_0(\vec{r};z_1,z_2)\equiv
V_0^{(1)}(|\vec{r}-z_1\vec{e}_z|) 
+V_0^{(2)}(|\vec{r}-z_2\vec{e}_z|),
\end{equation}
where we take only the central part of the potential \cite{Diaz-Torres05}. In Eq. (\ref{eq:volume}), 
$\theta$ is a step function and 
$W_0$ is a constant, 
for which we take $-12$ MeV so that the value of $W_0$ is around the Fermi energy for the $^{32}$S nucleus. 
Notice that, in contrast to Refs. \cite{Diaz-Torres05,NSP87}, we take into account 
in Eq. (\ref{eq:volume}) 
the effect of 
finite surface diffuseness by evaluating the volume integral of the 
potential.  
We find that this is important in order to 
keep the depth parameter in a Woods-Saxon potential 
similarly to each other between $^{16}$O and $^{32}$S. 
Notice also that we interpolate the inverse of the radius parameter, $1/R$, rather 
than the radius parameter 
itself. We find that this scheme is more convenient for our purpose, as, 
at large distances $z$, 
the interpolation with $R$ with the volume conservation of Eq. (\ref{eq:volume}) 
tends to lead to a 
wider and shallower potential 
than the potential given by Eq. (\ref{eq:WS}), 
which however has the same volume integral 
to each other. 
This problem seems to disappear if the interpolation is carried out with the parameter $1/R$ rather than $R$. 

\begin{figure}[tb]
\includegraphics[clip,width=7cm]{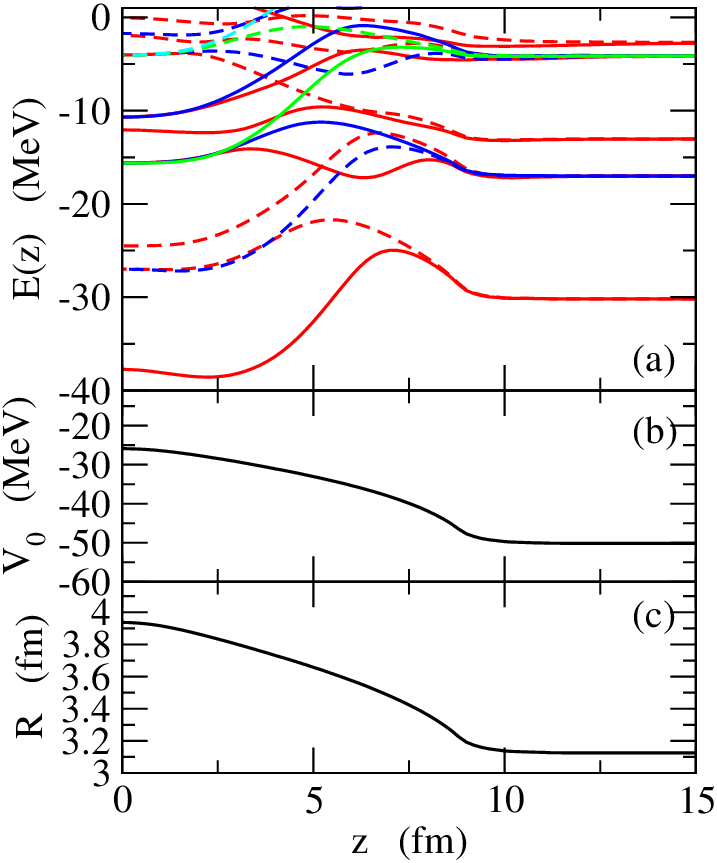}
\caption{(a) Neutron single-particle energies for the $^{16}$O+$^{16}$O 
system as a function of the separation distance between the two 
$^{16}$O nuclei. 
The solid and the dashed lines indicate the energies for the 
positive and the negative parity states, respectively. 
(b) The dependence of the depth parameter, $V_0$, in the Woods-Saxon potential on the 
separation distance as determined from the volume conservation condition 
given by Eq. (\ref{eq:volume}). 
(c) The same as the panel (b), but for the radius parameter, $R$. 
}
\end{figure}

The top panel of 
Fig. 2 shows the neutron single-particle energies so obtained 
as a function of 
the separation distance $z$ between the two potential wells, which are placed 
at $z_1=-z/2$ and $z_2=z/2$, respectively. 
The $z$-dependence of the depth and the radius parameters in 
the Woods-Saxon potentials are also shown in the middle and the bottom 
panels, respectively.
Since the two-center potential is axially symmetric around the $z$-axis, 
and also it is symmetric with respect to the parity transformation, the 
$z$-component of the total angular momentum, $j_z=\Lambda+m_s$, as well 
as the parity, $\pi$, 
are good quantum numbers to characterize the single-particle 
states. 
In the figure, the positive and negative parity states are indicated by the solid and 
the dashed lines, respectively. 
To obtain these energies, we expand each potential on the harmonic oscillator 
potential within 12 major shells, for which the frequency of the harmonic 
oscillator is taken to be $\hbar\omega_0=41\times 16^{-1/3}$ MeV. 
We have confirmed that the results do not change significantly even when 
a larger number of basis states are taken into account.
For the expansion of the single-particle wave functions, we include 
the eigen-functions given by Eq. (\ref{eigen-z}) up to $\epsilon_z=20$ MeV, 
where the continuum states are discretized with the box boundary 
condition, with the box size of $z_{\rm box}=z/2+10$ fm. 
As one can see, well-known features of single-particle energies in a 
symmetric two-center potential \cite{ZM76,MR86,GGR77} 
are well reproduced also in this calculation. 
That is, at large separation distances, positive and negative parity states 
are degenerate in energy, as they correspond to the symmetric 
and the anti-symmetric combinations of the wave function for the 
same state in the 
right and the left potential wells, respectively. As the separation distance 
decreases, these states are bifurcated, and 
the positive (negative) 
parity combination 
is converged to one of the positive-parity (negative-parity) 
single-particle states in the 
unified system in the limit of zero separation distance. 
At the intermediate separation distances, one can see a few avoided 
level crossings in a pair of single-particle states with the same parity and $j_z$.  

\begin{figure}[tb]
\includegraphics[clip,width=7cm]{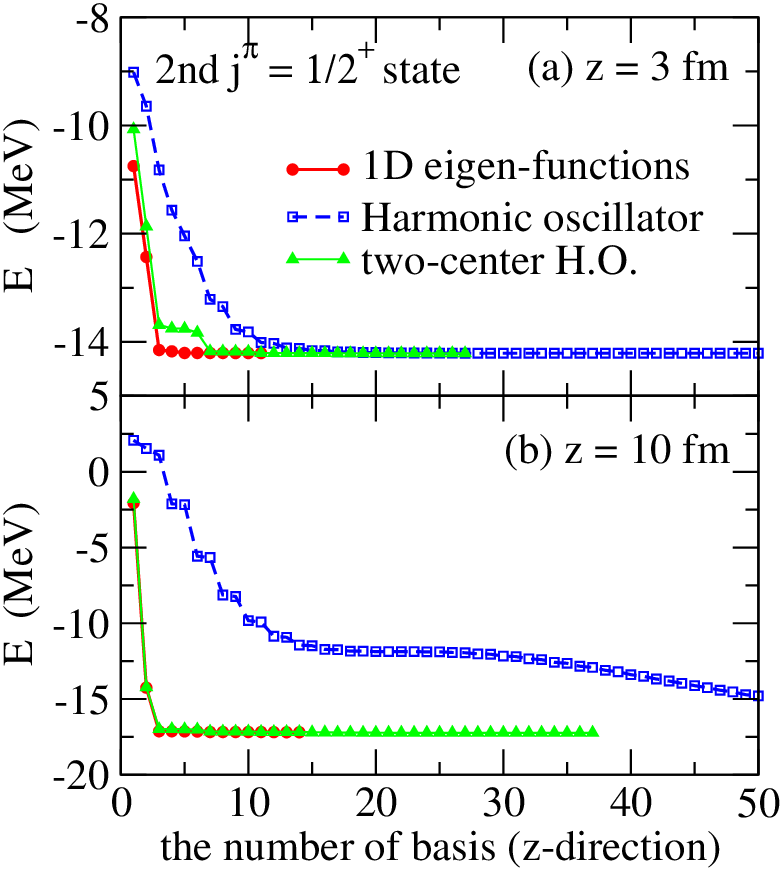}
\caption{
The single-particle energy for the second positive-parity state with 
$j_z^\pi=1/2^+$ 
as a function of the number of basis states  in Eq. (\ref{hobasis_wf}) 
for the $z$-direction. 
The upper and the lower panes show the result at the 
separation distance of $z=3$ fm and $z=10$ fm, respectively. 
The solid lines with filled circles indicate the results with the 
eigen-solutions for the one-dimensional two-center potentials 
given by Eq. (\ref{eigen-z}), while the dashed lines and the solid lines with filled triangles 
show the results with one-center and two-center harmonic oscillator bases, respectively. }
\end{figure}

\begin{figure}[tb]
\includegraphics[clip,width=7cm]{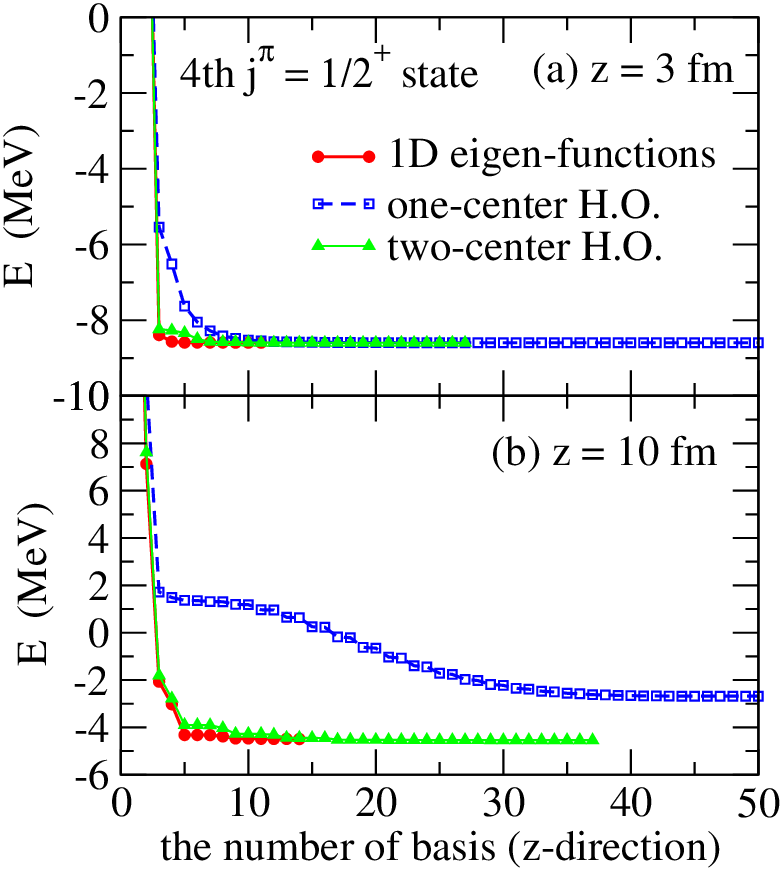}
\caption{
Same as Fig. 3, but for the 4th positive-parity state with $j_z^\pi=1/2^+$.}
\end{figure}

The convergence feature of the calculation is shown by the 
solid line with filled circles in Figs. 3 and 4. These are the 
single-particle energies for the second and the fourth 
positive-parity states with $j_z^\pi=1/2^+$ (see Fig. 2(a)), respectively,
at $z=3$ fm (the upper panel) and at 
$z=10$ fm (the lower panel) as a function 
of the number of the basis states in 
Eq. (\ref{hobasis_wf}) for the $z$-direction. 
For a comparison, we also show 
the results with a one-dimensional harmonic 
oscillator basis with a single center (the dashed line) and a double center (the solid line 
with filled triangles). The former is obtained 
with the oscillator length of 
$b_z={\rm max}(z,b_0)$, where ${\rm max}(a,b)= a$ for $a\geq b$ and 
${\rm max}(a,b)= b$ for $a< b$, while the latter uses the same oscillator length as the one used to expand the 
potential. Notice that the latter basis can be constructed analytically 
with confluent hypergeometric functions \cite{HMG69}. 
As expected, the convergence is 
fast for this calculation, 
while a similar good convergence is achieved also 
with a two-center harmonic oscillator basis.
That is, the energy is almost converged by including 
only a few eigen-functions in the $z$-direction, both at $z=3$ fm and $z=10$ fm. 
In contrast, the convergence is considerably slow with the single-center harmonic 
oscillator basis, especially at large separation distances. 
Evidently, the method proposed in this paper provides a powerful 
way to solve the two-center shell model with arbitrary finite 
depth potential wells. 

\section{Summary}

We have proposed a novel method to solve the eigen-value 
problem for a single-particle motion in a two-center potential. 
This method combines the separable representation for the single-particle 
potential with the usual matrix diagonalization. 
To this end, we have expanded the potential on a harmonic oscillator 
basis while the single-particle wave functions on the combined basis 
of a harmonic oscillator and eigen-functions of a one-dimensional 
two-center potential. 
In this way, the method can be applied easily and efficiently even 
to a situation with two 
close single-particle energies. Also, with this method both 
bound and resonance states 
can be obtained in a single framework. 

In this paper, we have considered a two-center potential which consists of 
two shifted spherical Woods-Saxon 
potentials. It would be an interesting future 
problem to 
extend the present approach to a two-center potential with deformed 
potentials \cite{NSP87,Diaz-Torres08}. Such extension would 
be useful in order to understand the reaction dynamics for hot fusion reactions 
to synthesize superheavy elements, in which $^{48}$Ca beams are used 
together with a deformed actinide target nucleus \cite{OU15,HHO13}. 

\acknowledgments

This work was supported by JSPS KAKENHI Grant Number 
15K05078. 

\appendix

\section{Matrix elements of Hamiltonian with the harmonic oscillator basis}

In this Appendix, we present matrix elements of the Hamiltonian 
with a spherical potential, Eq. (\ref{Vsphe}), with the harmonic oscillator 
basis, Eq. (\ref{hobasis}), in the cylindrical coordinate. 
To this end, we closely follow Sec. IV-C in Ref. \cite{Vautherin73}. 

\subsection{Central potential}

We first consider the central part of the potential, 
$V_0^{(s)}(r)$. 
Its matrix elements read, 
\begin{eqnarray}
\langle \Psi_\alpha|V_0^{(s)}|\Psi_{\alpha'}\rangle 
&=& \delta_{\Lambda,\Lambda'}\delta_{m_s,m_s'} \nonumber \\
&&\times\int^\infty_{-\infty}dz \int^{\infty}_0\rho d\rho\,
V_0^{(s)}(r)\,
\phi_{n_z}(z)\phi_{n_z'}(z) \nonumber \\
&&\times \psi_{n_\rho}^{(\Lambda)}(\rho)\psi_{n_\rho'}^{(\Lambda')}(\rho),
\end{eqnarray}
with $r=\sqrt{\rho^2+z^2}$. 

\subsection{Spin-orbit potential}

We next consider the spin-orbit potential, 
$V_{\rm ls}^{(s)}(r)\,\vec{l}\cdot\vec{s}$. 
We first note that the basis function 
$|\Psi_{\alpha}\rangle$ is an eigen-function of $l_z$ and $s_z$ as, 
\begin{eqnarray}
l_z\Psi_\alpha &=& \frac{1}{i}\,\frac{\partial}{\partial\phi}\,\Psi_\alpha
= \Lambda\Psi_\alpha, \\
s_z\Psi_\alpha &=& m_s\,\Psi_\alpha. 
\end{eqnarray}
We also notice that 
\begin{equation}
\vec{l}\cdot\vec{s}=\frac{1}{2}(l_+s_- + l_-s_+) + l_zs_z,
\end{equation}
with $s_\pm=s_x\pm i s_y$ and 
\begin{eqnarray}
l_\pm & = & l_x \pm il_y, \\
&=&\mp e^{\pm i\phi}
\left(\rho\frac{\partial}{\partial z}-z\frac{\partial}{\partial \rho}
\pm \frac{z}{\rho}\,\frac{1}{i}\frac{\partial}{\partial \phi}\right). 
\end{eqnarray}
Since $s_+\chi_{\downarrow}=\chi_\uparrow$, $s_-\chi_{\uparrow}=\chi_\downarrow$, 
and
\begin{equation}
\int^{2\pi}_0\frac{e^{-i\Lambda\phi}}{\sqrt{2\pi}}\,
(\mp e^{\pm i\phi})\,\frac{e^{i\Lambda'\phi}}{\sqrt{2\pi}}
=\mp \delta_{\Lambda,\Lambda'\pm1}, 
\end{equation}
one obtains, 
\begin{eqnarray}
&&\langle\Psi_{n_zn_\rho\Lambda m_s}
|V_{\rm ls}^{(s)}(r)\vec{l}\cdot\vec{s}|\Psi_{n'_zn'_\rho\Lambda' m_s}\rangle 
\nonumber \\
&&\quad=\delta_{\Lambda,\Lambda'} \nonumber \\
&&\quad\times\Lambda m_s\,\int^\infty_{-\infty}dz \int^{\infty}_0\rho d\rho\,
V_{\rm ls}^{(s)}(r)\,
\phi_{n_z}(z)\phi_{n_z'}(z) \nonumber \\
&&\quad\times \psi_{n_\rho}^{(\Lambda)}(\rho)\psi_{n_\rho'}^{(\Lambda')}(\rho), \\
&&\langle\Psi_{n_zn_\rho\Lambda \downarrow}
|V_{\rm ls}^{(s)}(r)\vec{l}\cdot\vec{s}|\Psi_{n'_zn'_\rho\Lambda' \uparrow}\rangle 
\nonumber \\
&&\quad=-\frac{1}{2}\,\delta_{\Lambda,\Lambda'+1} \nonumber \\
&&\quad\times\int^\infty_{-\infty}dz \int^{\infty}_0\rho d\rho\,
V_{\rm ls}^{(s)}(r)\,
\phi_{n_z}(z)\psi_{n_\rho}^{(\Lambda)}(\rho)\nonumber \\
&&\quad\times 
\left(\rho\frac{\partial}{\partial z}-z\frac{\partial}{\partial \rho}
+ \frac{z}{\rho}\,\Lambda'\right) 
\phi_{n_z'}(z) 
\psi_{n_\rho'}^{(\Lambda')}(\rho),
\end{eqnarray}
and
\begin{eqnarray}
&&\langle\Psi_{n_zn_\rho\Lambda \uparrow}
|V_{\rm ls}^{(s)}(r)\vec{l}\cdot\vec{s}|\Psi_{n'_zn'_\rho\Lambda' \downarrow}\rangle 
\nonumber \\
&&\quad=\frac{1}{2}\,\delta_{\Lambda,\Lambda'-1} \nonumber \\
&&\quad\times\int^\infty_{-\infty}dz \int^{\infty}_0\rho d\rho\,
V_{\rm ls}^{(s)}(r)\,
\phi_{n_z}(z)\psi_{n_\rho}^{(\Lambda)}(\rho)\nonumber \\
&&\quad\times 
\left(\rho\frac{\partial}{\partial z}-z\frac{\partial}{\partial \rho}
- \frac{z}{\rho}\,\Lambda'\right) 
\phi_{n_z'}(z) 
\psi_{n_\rho'}^{(\Lambda')}(\rho). 
\end{eqnarray}

\subsection{Kinetic energy}

The matrix elements for the kinetic energy for the $z$-direction 
is computed as 
\begin{eqnarray}
\left\langle \Psi_\alpha\left|\frac{p_z^2}{2m}\right|\Psi_{\alpha'}\right\rangle 
&=& -\frac{\hbar^2}{2m}\,\delta_{n_\rho,n'_\rho}\delta_{\Lambda,\Lambda'}\delta_{m_s,m_s'} \nonumber \\
&&\times\int^\infty_{-\infty}dz \,
\phi_{n_z}(z)
\frac{d^2\phi_{n_z'}(z)}{dz^2}. \nonumber \\
&=& \frac{\hbar^2}{2m}\,\delta_{n_\rho,n'_\rho}\delta_{\Lambda,\Lambda'}\delta_{m_s,m_s'} \nonumber \\
&&\times\int^\infty_{-\infty}dz \,
\frac{d\phi_{n_z}(z)}{dz}
\frac{d\phi_{n_z'}(z)}{dz}. \nonumber \\
\end{eqnarray}
Here, we keep the matrix elements in a general form, so that the 
formula can be applied also to Eq. (\ref{T}) with Eq. (\ref{hobasis_wf}). 

In order to evaluate the matrix elements for the kinetic energy for 
the $\rho$-direction, we use the Schr\"odinger equation of, 
\begin{equation}
\left(\frac{p_x^2+p_y^2}{2m}+\frac{1}{2}m\omega_0^2\rho^2-\epsilon_{n_\rho\Lambda}
\right)\Psi_\alpha(\vec{r})=0,
\end{equation}
with $\epsilon_{n_\rho \Lambda}=(2n_\rho + \Lambda+1)\,\hbar\omega_0$. 
This leads to, 
\begin{eqnarray}
&&\left\langle \Psi_\alpha\left|\frac{p_x^2+p_y^2}{2m}
\right|\Psi_{\alpha'}\right\rangle \nonumber \\
&&= \epsilon_{n_\rho \Lambda}\,\delta_{\alpha,\alpha'} \nonumber \\
&&
-\delta_{n_z,n'_z}\delta_{\Lambda,\Lambda'}\delta_{m_s,m_s'}\,
 \nonumber \\
&&\quad\times \int^{\infty}_0\rho d\rho\,
\left(\frac{1}{2}m\omega_0^2\rho^2\right)\psi_{n_\rho}^{(\Lambda)}(\rho)\psi_{n_\rho'}^{(\Lambda')}(\rho). 
\end{eqnarray}

\end{document}